\documentclass[12pt,aps,prd,showpacs,nofootinbib,preprintnumbers,a4paper]{revtex4-1}

\usepackage{amsmath}
\usepackage{comment}
\usepackage[dvipdfmx]{graphicx}

\usepackage{dcolumn}% Align table columns on decimal point
\usepackage{bm}% bold math
\usepackage[dvipdfmx]{color}
\usepackage[normalem]{ulem}
\allowdisplaybreaks

\newcommand{\nn}{\nonumber}

\newcommand{\be}{\begin{eqnarray}}
\newcommand{\ee}{\end{eqnarray}}
\catcode`\@=11
\def\lsim{\mathrel{\mathpalette\@versim<}}
\def\gsim{\mathrel{\mathpalette\@versim>}}
\def\@versim#1#2{\vcenter{\offinterlineskip
\ialign{$\m@th#1\hfil##\hfil$\crcr#2\crcr\sim\crcr } }}
\catcode`\@=12
\newcommand{\del}{\partial}

\newcommand{\Slash}[1]{{\ooalign{\hfil#1\hfil\crcr\raise.167ex\hbo x{/}}}}
\newcommand\normalorder[1]{{:}\mkern1mu#1\mkern1.6mu{:}}
\def\thefootnote{\fnsymbol{footnote}}

\begin{document}

\title{
 Genesis of electroweak and dark matter scales from a bilinear scalar
    condensate}

\author{Jisuke \surname{Kubo}}
\email{jik@hep.s.kanazawa-u.ac.jp}
\affiliation{Institute for Theoretical Physics, Kanazawa University, Kanazawa 920-1192, Japan}

\author{Masatoshi \surname{Yamada}}
\email{masay@hep.s.kanazawa-u.ac.jp}
\affiliation{Institute for Theoretical Physics, Kanazawa University, Kanazawa 920-1192, Japan}

\preprint{KANAZAWA-15-05}


\begin{abstract}
The condensation of scalar bilinear in a classically scale invariant strongly interacting
hidden sector is used to generate the electroweak scale, where the excitation of the condensate is identified as dark matter.
We formulate an effective theory for the condensation of the scalar bilinear and find in the self-consistent mean field approximation  that 
the dark matter mass is of $O(1)$ TeV with 
the spin-independent elastic cross section off the nucleon 
slightly below the LUX upper bound.

 \end{abstract}
\setcounter{footnote}{0}
\def\thefootnote{\arabic{footnote}}
\maketitle

\section{Introduction}
Can we explain the origin of ``mass without mass'' \cite{wilczek}?
Yes,  a large portion  of  the baryon mass 
can be produced by 
dynamical chiral symmetry breaking  (D$\chi$SB) ``from nothing'' 
\cite{Nambu:1960xd,Nambu:1961tp}.
This nonperturbative mechanism, instead of
the Brout-Englert-Higgs mechanism,  can also be applied to trigger electroweak symmetry breaking  \cite{Weinberg:1975gm,Susskind:1978ms}. 
After the discovery of the Higgs particle \cite{Aad:2012tfa,Chatrchyan:2012ufa},
however, it is a fair assumption that  fundamental scalars can exist.
Since the Higgs mass term is the only term in the standard model (SM),
that breaks scale invariance at the classical level,
we can thus  ask where the Higgs mass term comes from.
Even the Higgs mass term, too, may have its origin in 
a nonperturbative  effect.
In fact D$\chi$SB
 in  a QCD-like hidden sector
has been recently used to induce the  Higgs mass term in a classically scale invariant extension of the SM 
\cite{Hur:2007uz,Heikinheimo:2013fta,Holthausen:2013ota,
Kubo:2014ida}.

In this paper we focus on another nonperturbative effect,
the condensation of the scalar bilinear (CSB)
\cite{Abbott:1981re,Chetyrkin:1982au} 
(see also \cite{Osterwalder:1977pc,Fradkin:1978dv}) in a strongly interacting hidden sector, 
to generate directly the Higgs mass term via the Higgs portal 
\cite{Kubo:2014ova}.
The main difference between two classes of models, apart from
how a scale is dynamically generated, 
is that in the first class of models (with D$\chi$SB)
the scale generated  in a hidden sector has to be transmitted to the SM via a
mediator, e.g. a SM singlet scalar in the model considered in 
\cite{Hur:2007uz,Heikinheimo:2013fta,Holthausen:2013ota,
Kubo:2014ida},
while such a mediator is not needed in the second class of 
models (with CSB).
This will be an important difference if two classes should be experimentally 
distinguished.
Another important difference is that the DM particles of the first class
are $CP$-odd scalars, 
while they are $CP$-even scalars in the second class, as we will see.

Our interest in the second class of  models is twofold:
first, 
because the discussion on how the Higgs mass is generated in \cite{Kubo:2014ova} 
is rather qualitative, we here formulate
an effective theory to nonperturbative breaking of scale invariance
by the CSB. This enables us
to  perform an approximate but quantitative treatment.
Second, since only one flavor for the strongly interacting scalar field
$S$ is considered in \cite{Kubo:2014ova} so that there is no
dark matter (DM) candidate, we  introduce $N_f$ flavors and investigate
whether we can obtain   realistic candidates of DM.
The DM candidates in our scenario are scalar-antiscalar bound states,
which are introduced as the excitation of the condensate
 in the self-consistent
mean field approximation (SCMF) 
\cite{Kunihiro:1983ej,Hatsuda:1994pi}. Their interactions with the SM
can be obtained by integrating 
out the ``constituent'' scalars.
In this approximation we can constrain the parameter space
of the effective theory in which realistic DM candidates are  present.

\section{The model and its effective Lagrangian}
We start by considering a hidden sector described by an $SU(N_c)$ gauge 
theory with the scalar fields $S_i^{a}~
(a=1,\dots,N_c,~i=1,\dots,N_f)$ in
the fundamental representation of $SU(N_c)$.
The Lagrangian of the hidden sector
is 
\be
{\cal L}_{\rm H} &=&-\frac{1}{2}~\mbox{tr} F^2+
([D^\mu S_i]^\dag D_\mu S_i)-
\hat{\lambda}_{S}(S_i^\dag S_i) (S_j^\dag S_j)\nn\\
& &-\hat{\lambda'}_{S}
(S_i^\dag S_j)(S_j^\dag S_i)
+\hat{\lambda}_{HS}(S_i^\dag S_i)H^\dag H,
\label{LH}
\ee
where $D_\mu S_i = \partial_\mu S_i -ig_{\rm H} G_\mu S_i$, $G_\mu$ is the matrix-valued gauge field, 
the trace is taken over  the color indices,
and the parentheses in Eq.~\eqref{LH} stands for an $SU(N_c)$ invariant product.
The SM Higgs doublet field is denoted by $H$.
The total Lagrangian is
${\cal L}_T ={\cal L}_{\rm H}+{\cal L}_{\mathrm{SM}} $, 
and the SM part, ${\cal L}_{\mathrm{SM}}$, contains
 the SM gauge and Yukawa  interactions
along with the scalar potential
$V_{\mathrm{SM}}
=\lambda_H ( H^\dag H)^2$
 without the Higgs mass term.

We assume that  for {a certain energy} the gauge coupling $g_{\rm H}$ becomes so strong that
the $SU(N_c)$ invariant scalar bilinear
forms  a $U(N_f)$ invariant
condensate \cite{Abbott:1981re,Chetyrkin:1982au}
\be
\langle (S^\dag_i S_j)\rangle &=&
\langle ~\sum_{a=1}^{N_c} S^{a\dag}_i S^a_j~\rangle\propto \delta_{ij}.
\label{condensate}
\ee 
This  nonperturbative condensate breaks scale invariance, but it is not an order parameter, because scale invariance is broken by scale anomaly \cite{Callan:1970yg}.
The breaking by anomaly is hard but only logarithmic, which means basically that the coupling constants depend on the energy scale~\cite{Callan:1970yg}.
{Moreover, we should note that the mass term is not generated by the anomaly since the beta function of the mass is propotional to the mass itself, see e.g.~\cite{Ford:1992mv}.\footnote{
{
In viewpoint of Wilsonian renormalization group, the classical scale invariance means that the bare mass is exactly put on the critical surface~\cite{Aoki:2012xs}.
Once this tuning is done, the renormalized mass keeps vinishing under the renoramlization group transformation.
}
}
The creation of the mass term from nothing can happen only by a nonperturbative effective, i.e. the condensate \eqref{condensate} is taken place.\footnote{
{
Once the mass is dynamically generated, the scale anomaly contributes to the mass.
}
}}
Therefore, the non-perturbative  breaking due to the  condensation may be assumed to be dominant, so that we can ignore the breaking by scale anomaly in the lowest order approximation to the breaking of scale invariance.

Under this assumption the condensate 
is a good order parameter, and we would like to 
formulate an effective theory,
which is an analog of the Nambu--Jona-Lasinio (NJL) theory \cite{Nambu:1961tp} to
D$\chi$SB.
The Lagrangian  of the effective theory
will not contain the $SU(N_c)$ gauge fields, because 
they are integrated out, while it contains
the ``constituent'' scalar fields $S_i^{a}$, for which 
{\em we use the same symbol as the original scalar fields}.
Since the effective theory
should  describe the symmetry breaking dynamically,
the effective Lagrangian has to be invariant under
 the  symmetry transformation in question:
 \be
 {\cal L}_{\rm eff} &=& 
 ([\partial^\mu S_i]^\dag \partial_\mu S_i)-
\lambda_{S}(S_i^\dag S_i) (S_j^\dag S_j)
-\lambda'_{S}
(S_i^\dag S_j)(S_j^\dag S_i)\nn\\
& &+\lambda_{HS}(S_i^\dag S_i)H^\dag H
-\lambda_H ( H^\dag H)^2,
\label{Leff}
\ee
with all positive $\lambda$'s. 
This is the most general form which is
consistent with 
the $SU(N_c)\times U(N_f)$ symmetry and the classical scale invariance,
where we have not included the kinetic term for $H$ 
in  ${\cal L}_{\rm eff}$, because it does not play
any  significant role as far as the effective theory for
the CSB is concerned.\footnote{Quantum field theory defined by (\ref{Leff})
with the kinetic term for $H$ is renormalizable 
in perturbation theory \cite{Lowenstein:1975rf}.}
Note that the couplings
$\hat{\lambda}_{S}$, $\hat{\lambda'}_{S}$ and $\hat{\lambda}_{HS}$ in ${\cal L}_{\rm H}$ of (\ref{LH}) are not the same as
$\lambda_{S}$,  $\lambda'_{S}$ and $\lambda_{HS}$ in $ {\cal L}_{\rm eff}$,
 respectively. 
{We emphasize that the effective Lagrangian \eqref{Leff} is scaleless, and describes the dynamics of scalar field $S$ at slightly above the confinement scale, thus,  the scalar condensate has not taken place yet.
Therefore the mixing of multiple scales discussed in \cite{Aoki:2012xs} does not appear.\footnote{{
Although NJL model is also defined before the dynamical breaking of chiral symmetry, its Lagrangian has a scale at which the Lagrangian is given.}
}
Using the effective Lagrangian \eqref{Leff}, we attempt to approximately describe the genesis of scale by the original gauge theory \eqref{LH} as the ``non-perturbative'' dimensional transmutation, {\it \`{a}~la} Coleman--Weinberg.
In the following, we demonstrate this mechanism and present our formalism by considering first $N_f=1$ case.
}

\vspace{0.2cm}
\noindent
A. {\bf $N_f=1$ (with $\lambda'_{S}=0)$}\\
In the SCMF approximation,
which has proved 
to be a successful approximation for  the NJL theory
\cite{Kunihiro:1983ej},
the perturbative vacuum  is 
Bogoliubov-Valatin (BV) transformed to $ | 0_{\rm B} \rangle$, such that
 $\langle 0_{\rm B}  | (S^\dag S)| 0_{\rm B}  \rangle =
f$, where $f$ has to be determined in a self-consistent way.
One  first splits up the effective Lagrangian (\ref{Leff}) into the sum
$\mathcal{L}_{\rm eff} =\mathcal{L}_{\rm MFA}+\mathcal{L}_{I}$,
where  $\mathcal{L}_{I}$ is normal ordered 
(i.e. $\langle 0_{\rm B}\vert \mathcal{L}_{I}\vert 0_{\rm B}\rangle =0$), and $\mathcal{L}_{\rm MFA}$ contains at most the bilinears of $S$ 
 which are not normal ordered. Using the Wick theorem
$(S^\dag S) =\normalorder{(S^\dag S)} +f$,
$(S^\dag S)^2 =\normalorder{(S^\dag S)^2}+2f(S^\dag S)-f^2$,
etc., we find
\be
\mathcal{L}_{\rm MFA}
&=&(\del^\mu S^\dag\del_\mu S)
-M^2(S ^\dagger S)
-\lambda_H(H^\dagger H)^2+\lambda_Sf^2,\nn
\ee
where 
$M^2= 2\lambda_S f-\lambda_{HS} H^\dag H$.
To the lowest order in the SCMF approximation
the ``interacting '' part $\mathcal{L}_I$ does not contribute
to the amplitudes without external $S$s (the mean field vacuum amplitudes).
We emphasize that, in applying  the Wick theorem,  
 only the $SU(N_c)$ invariant bilinear product
$(S^{\dag} S)=\sum_a^{N_c} S^{a\dag} S^a$
 has a nonzero  (BV transformed) vacuum expectation value.
To compute loop corrections we employ the $\overline{\mbox{MS}}$
scheme, 
because dimensional regularization does not break scale invariance.
To the lowest order the divergences can be removed by
renormalization of $\lambda_I~(I=H,S,HS)$, i.e.
$\lambda_I\to (\mu^2)^\epsilon (\lambda_{I}+\delta\lambda_{I})$,
and also by  the shift
$f\to f+\delta f$,
where $\epsilon =(4-D)/2$, and $\mu$ is the  scale introduced in dimensional regularization.
The effective potential for $\mathcal{L}_{\rm MFA}$ can be straightforwardly 
computed :
\be
V_{\rm MFA}
&=&
M^2(S ^\dagger S)
+\lambda_H(H^\dagger H)^2-
\lambda_Sf^2+\frac{N_c}{32\pi^2}
M^4\ln\frac{M^2}{\Lambda_H^2},
\label{VMFA}
\ee
where $\Lambda_H=\mu \exp (3/4)$ is chosen such that the loop correction vanishes at
$M^2=\Lambda_H^2$.
($V_{\rm MFA}$ with a term linear in $f$ included but without the Higgs doublet $H$ has been discussed in 
\cite{Coleman:1974jh,Kobayashi:1975ev,Abbott:1975bn}.
The classical scale invariance forbids the presence of this linear term.)
{
Note here that the scale $\Lambda_H$ is generated by the non-perturbative loop effect.}
To find the minimum of 
$V_{\rm MFA}$ we look for the solutions of
\be
0&=&\frac{\del}{\del S^a}V_{\rm MFA}
= \frac{\del}{\del f}V_{\rm MFA}
=\frac{\del}{\del H_l}V_{\rm MFA}~(l=1,2).
\label{station}
\ee
The first equation gives
$0=(S^{a})^{\dag}M^2=(S^{a})^{\dag}( 2\lambda_S f-\lambda_{HS} H^\dag H)$,
which has three solutions: (i) $ \langle S^{a} \rangle \neq 0~\mbox{and}~\langle M^ 2\rangle =0$, 
(ii) $ \langle S^{a} \rangle = 0~\mbox{and}~\langle M^2 \rangle =0$, and (iii)
$ \langle S^{a} \rangle = 0~\mbox{and}~\langle M^2 \rangle\neq 0$.
%%%%
The effective potential $V_{\rm MFA}$ in the solution
(i) has a flat direction, which corresponds to the end-point contribution discussed 
in~\cite{Bardeen:1983st}.
In the flat direction
 (i.e. $ f=
 H=0$), $V_{\rm MFA}=0$ for any value of 
$S^a$, so that the $SU(N_{\rm c})$ symmetry is spontaneously broken.
If all the extremum conditions (\ref{station}) are imposed
for the solution (i), we obtain
$\langle f\rangle=\langle ([S^{a}]^\dag S^{a})\rangle
=(2\lambda_{H}/\lambda_{HS}) \langle H^\dag H\rangle$
along with $C=0$ and $\langle V_{\rm MFA}\rangle=0$ \cite{Coleman:1974jh}.\footnote{
Due to $\langle M^2 \rangle =0$ there exists a tachyonic state,
because the inequality of  \cite{Kobayashi:1975ev},
$16 \pi^2/(2 N_c \lambda_S) 
-\ln [\langle M^2 \rangle/\Lambda_H^2 \exp (-3/2)] < 0$,
cannot be satisfied for a finite $\Lambda_H$ and a positive $\lambda_S$.
}
%%%%
%Therefore, the solution (i) is inconsistent,
%unless we use the fine-tuned relation among the coupling constants.
Next we consider (ii) and find that $ \langle S^{a}
 \rangle =\langle f \rangle=\langle H\rangle =0$
 with  $\langle V_{\rm MFA}\rangle =0$.
The third solution  (iii) can exist
if
\be
C&=& 4\lambda_H \lambda_S-\lambda_{HS}^2 > 0 
\label{C}
 \ee
 is satisfied, and 
 we find
 \be
 |\langle H\rangle |^2
 &=&v_h^2/2=
 \frac{\lambda_{HS}}{C}\Lambda_H^2\exp\left(  \frac{32\pi^2 \lambda_H}{N_c C}-\frac{1}{2}\right)~,~
 \langle f\rangle =\frac{2 \lambda_H}{\lambda_{HS}} 
 |\langle H\rangle |^2 ,\nn\\
 \langle V_{\rm MFA} \rangle &=& -
 \frac{N_c }{64\pi^2} \Lambda_H^4
\exp\left(  \frac{64\pi^2 \lambda_H}{N_c C}-1\right) < 0.\nn
 \ee
Consequently, the solution  (iii)
presents  the true potential minimum  
if (\ref{C})  is satisfied (in the energy region 
where (\ref{Leff}) should serve as the effective Lagrangian).
Self-consistency means that 
$f=\langle 0_{\rm B}  | (S^\dag S)| 0_{\rm B}  \rangle$
is equal to $\langle f \rangle$ at  the potential minimum 
in the mean field  approximation.
 The Higgs mass at this level of approximation becomes
\be
m_{h0}^2 
&=&\frac{\lambda_{HS}\Lambda_H^2}{C}\left(
\frac{16 \lambda_H^2\lambda_S}{C}
+\frac{N_c \lambda_{HS}^2}{8\pi^2}
\right)
 \exp\left(  \frac{32\pi^2 \lambda_H}{N_c C}-\frac{1}{2}\right).
 \label{mh0}
\ee
In the  small $\lambda_{HS}$ limit we obtain
$m_{h0}^2 \simeq 4 \lambda_{H}| \langle H\rangle |^2
=2 \lambda_{HS}\langle f \rangle$, where
the first equation is the SM expression, 
and the second one is simply assumed in \cite{Kubo:2014ova}.
So the Higgs mass (\ref{mh0}) contains the backreaction.
The analysis above shows that the scale created in the hidden sector 
can be desirably transmitted to the SM sector.
The reason that $\langle V_{\rm MFA} \rangle  <0$
for the solution (iii) is the absence of a mass term in the effective Lagrangian
(\ref{Leff});  the classical scale invariance does not allow the mass term.
A mass term in (\ref{Leff}) 
would  generate a term linear in $f$  in 
$V_{\rm MFA}$, which can lift the 
$\langle V_{\rm MFA} \rangle$ into a positive direction
\cite{Kobayashi:1975ev,Abbott:1975bn}, while
$V_{\rm MFA}=0$ remains in  the flat direction  \cite{Bardeen:1983st}.

At this stage we would like to mention that Bardeen and Moshe 
\cite{Bardeen:1983st} (and also others)
pointed out the intrinsic instability  inherent in  (\ref{Leff})
 (which is related to  its triviality) if one regards (\ref{Leff}) as a fundamental Lagrangian.
We however discard this fundamental problem,
because we assume that such a problem is absent in 
the original theory described by (\ref{LH}).

\vspace{0.2cm}
\noindent
B. { \bf $N_f> 1 $ and dark matter}\\
Here we consider the case with $N_f> 1 $ and 
take into account the excitations of the condensate, $\sigma$ and $\phi^\alpha~
(\alpha=1,\dots,N_f^2-1)$, which are introduced as
\be
\langle 0_{\rm B}  | (S_i^\dag S_j)| 0_{\rm B}  \rangle &=&
f_{ij}=\langle f_{ij}\rangle +Z_{\sigma}^{1/2}\delta_{ij}\sigma +
Z_{\phi}^{1/2}t_{ji}^\alpha \phi^\alpha.\label{condens}
\ee
Here $t^\alpha$ are
the $SU(N_f)$ generators in the Hermitian matrix representation, and
$Z_\sigma$ and $Z_\phi$ are the  wave function renormalization constants
of  a canonical  dimension $2$.
The unbroken $U(N_f)$ flavor symmetry  implies
 $\langle f_{ij}\rangle =\delta_{ij} f_0$ and 
 $\langle \phi^\alpha\rangle =0$, where $\langle \sigma\rangle$
 can be  absorbed into $f_0$, so that we can always assume
 $\langle \sigma\rangle =0$.
Furthermore,  the flavor symmetry ensures the stability of $\phi^\alpha$, i.e.
they can be good DM candidates, because they are electrically neutral
and their interactions with the SM sector are loop suppressed,
as we will see.
%%%%%
%%%%%
Note that  $\sigma$ and $\phi^\alpha$ in (\ref{condens}) are introduced as c-numbers
without kinetic terms.
However, their kinetic terms  will be generated through $S^a$ loop effects, 
and consequently 
we will reinterpret them as quantum fields describing physical degrees of freedom.
%%%%%
%%%%%
The investigation of the vacuum structure is basically the same
as in the $N_f=1$ case. We are interested in 
the solution of type (iii) of the previous case, i.e.
$f_0\neq 0, |\langle H\rangle |=v_h/2\neq 0$, 
which is the true potential minimum  if
\be
G &=& 4N_f \lambda_H \lambda_S-N_f \lambda_{HS}^2+
4 \lambda_H\lambda'_S  >0
\ee
 is satisfied.
Similar calculations as in the previous case yield among other things
\be
m_{h0}^2 
&=&\frac{\lambda_{HS} N_f\Lambda_H^2}{G}\left(
\frac{16\lambda_H^2
(N_f\lambda_S+\lambda'_S)}{G}
+
\frac{N_c N_f
\lambda_{HS}^2}{8\pi^2}\right)\nn\\
& &\times
 \exp\left(  \frac{32\pi^2 \lambda_H}{N_c G}-\frac{1}{2}\right).
\label{vev2}
 \ee
The SCMF Lagrangian  $\mathcal{L}'_{\rm MFA}$ involving 
$\sigma$ and $\phi^\alpha$ can now be written as
 \be
\mathcal{L}'_{\rm MFA}
&=&(\del^\mu S_i^\dag\del_\mu S_i)
-M_0^2(S_i ^\dagger S_i)\nn\\
&+ &N_f(N_f\lambda_S+\lambda'_{S})Z_{\sigma}\sigma^2+
\frac{\lambda'_{S} }{2}Z_{\phi}\phi^\alpha \phi^\alpha \label{LMFAP}\\
& -&2(N_f \lambda_S+\lambda'_{S})
Z_{\sigma}^{1/2}\sigma(S_i ^\dagger S_i)
-2\lambda'_{S}Z_{\phi}^{1/2} (S_i^\dag t^\alpha_{ij} \phi^\alpha S_j)
\nn\\
&+ &\frac{\lambda_{HS}}{2}(S ^\dagger S)h
(2 v_h+h) -\frac{\lambda_H}{4}h^2
(6v_h^2+4 v_h h+h^2),\nn
\ee
where 
$M_0^2=2( N_f \lambda_S +\lambda'_{S})f_0
-\lambda_{HS} v_{h}^2/2$,
and $\mbox{Tr}(t^\alpha t^\beta)=\delta^{\alpha\beta}/2$. 
Further, $h$ is the Higgs field contained in the Higgs doublet
 as $H^T=( H^+,(v_h+h+i\chi)/\sqrt{2}  )$,
where $H^+$ and $\chi$ are the would-be Nambu-Goldstone fields.
Linear terms in $\sigma$ and $h$ are suppressed in 
(\ref{LMFAP}), because they will be canceled against the corresponding
tadpole corrections.

Using (\ref{LMFAP}) and integrating out the constituent
scalars $S^a_i$, we can obtain effective 
 interactions among $\sigma,~\phi$ and the Higgs $h$.
We first compute their inverse propagators,
up to and including one-loop order, to obtain 
their masses and the wave function renormalization constants:
\be
\Gamma_\phi^{\alpha\beta}(p^2)
&=&Z_{\phi}\delta^{\alpha\beta}\lambda'_{S}
\Gamma_\phi(p^2)=
Z_{\phi}\delta^{\alpha\beta}\lambda'_{S}
\left[1+
2\lambda'_{S} N_c\Gamma(p^2)\right],
\label{gamma}\\
\Gamma_\sigma(p^2)
&=&2Z_{\sigma}N_f(N_f\lambda_{S}+\lambda'_{S})\left[1+
2 N_c(N_f\lambda_{S}+\lambda'_{S})\Gamma(p^2)\right],\nn\\
\Gamma_{h\sigma}(p^2)
&=&-2Z_{\sigma}^{1/2}
 v_h\lambda_{HS}(N_f\lambda_S+\lambda'_{S})
N_f N_c ~\Gamma(p^2),\nn\\
\Gamma_h(p^2)
&=& p^2-m_{h0}^2+(v_h \lambda_{HS})^2
N_f N_c ~(\Gamma(p^2)-\Gamma(0)),\nn
\ee
where $m_{h0}^2$ is given in (\ref{vev2}), the canonical  kinetic term 
for $H$ is
included, and
\be
\Gamma(p^2) &=&
-\frac{1}{16\pi^2}\int_0^1 dx
 \ln\left[\frac{-x(1-x) p^2+M_0^2}{\Lambda_H^2 \exp(-3/2)}\right].\nn
\ee
We have included neither the wave function renormalization constant
for $h$ (which is approximately equal to 1 within the approximation
here) nor the corrections to 
$\Gamma_h$ coming from the SM sector
(which will only slightly influence our result).

The  DM mass is the zero of the inverse propagator, i.e.
\begin{eqnarray}
\Gamma_{\phi}^{\alpha\beta}(p^2 = {m_{\rm DM}}^2)=0~,
\label{zero}
\end{eqnarray}
and $Z_{\phi}$ 
(which has a canonical  dimension  $2$) can be obtained from
$Z_{\phi}^{-1} =\left.
2(\lambda'_{S})^2 N_c (d \Gamma/d p^2)\right|_{p^2=m_{\rm DM}^2}$.
The $\sigma$ and  Higgs  masses are obtained 
from the zero eigenvalues of the $h-\sigma$ mixing matrix.
Strictly speaking, this mixing has to be taken into account
in determining the renormalization constants
(matrix) for $\sigma$ and $h$. However, the mixing is
less than $1\%$ in a realistic parameter space so that
we ignore the mixing for the  renormalization constants. 
As we can see from (\ref{gamma}), the radiative correction to
the inverse propagator is proportional to $2\lambda'_{S} N_c/16 \pi^2$,
so that the solution of (\ref{zero}) for a real positive $p^2$ can exist if
$\lambda'_{S} N_c$ is sufficiently large. Therefore, if 
  an upper limit of $\lambda'_{S}$ is set, there will be a minimum value of $N_c$.
It turns out that
 the minimum $N_c$ is 3 for    $\Gamma_\phi(p^2)$ with $N_f=2$
 to have a zero if  $0 <  \lambda'_{S}< 2 \pi$.
 For a larger $N_f$ we need a larger $N_c$:
 the minimum  $N_c$ is 4 for $N_f=3$ for instance.
% \begin{center}
\begin{figure}
 \includegraphics[width=6cm]{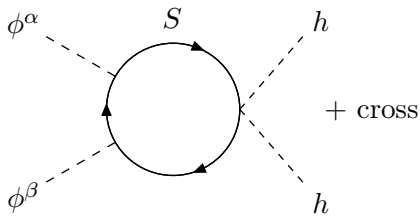}
\caption{\label{pp-hh}\footnotesize
The interaction between DM and the Higgs $h$
arises at the one-loop level.
Diagrams $\propto \lambda^2_{HS} (v_h/M_0)^2$
 are ignored, because 
 $\lambda^2_{HS} (v_h/M_0)^2 \ll \lambda_{HS}$.}
\end{figure}
%\end{center}
 
 The link of  $\phi$ to the SM model
 is established through the interaction with the Higgs,
 which is generated at one-loop as shown in Fig.~\ref{pp-hh}.
 %\footnote{There exist one-loop 
 %diagrams $\propto \lambda^2_{HS} (v_h/M_0)^2$
 %contributing to this interaction. We however ignored them, because 
% $\lambda^2_{HS} (v_h/M_0)^2 <<\lambda_{HS}$. }.
 We use the s-channel momenta $p=p'=(m_{\rm DM},{\bf 0})$ for
 DM annihilation,
because we restrict ourselves to the s-wave part of
the velocity-averaged annihilation cross section 
$\langle v\sigma \rangle$. 
For the spin-independent 
elastic cross section off the nucleon $\sigma_{SI}$
we use the t-channel momenta $p=-p'=(m_{\rm DM},{\bf 0})$. 
In these approximations the diagrams of 
Fig.~\ref{pp-hh} yield the effective couplings 
\be
\kappa_{s(t)} \delta^{\alpha\beta}
&=&  \delta^{\alpha\beta}\Gamma_{\phi^2 h^2}(M_0,m_{\rm DM},\epsilon=1(-1))~,
\label{kappa}
\ee
where
\be
\Gamma_{\phi^2 h^2}(M_0,m_{\rm DM}, \epsilon)
=\frac{Z_\phi  N_c (\lambda'_{S})^2\lambda_{HS}}{4\pi^2 }
\int_0^1 dx \int_0^{1-x} dy\left[M_0^2+m_{\rm DM}^2
(x(x-1)+y(y-1)-2 \epsilon xy)\right]^{-1},\nn
\ee
and we consider only the parameter space with
 $m_{\rm DM},\, m_\sigma < 2M_0$, because beyond 
that our SCMF approximation will  break down.
Then we obtain
\be
\langle v \sigma\rangle
&=&\frac{{1}}{32\pi m_{\rm DM}^3}~\sum_{I=W,Z,t,h}
(m_{\rm DM}^2-m_I^2)^{1/2} a_I+\mathcal{O}(v^2),\nn
\ee
where $m_{W,Z,t,h}$ are the $W, Z$, top quark
and Higgs masses, respectively, and
\be
& &a_{W(Z)}= 4 (2)\kappa_s^2\Delta_{h}^2 m_{W(Z)}^4
\left( 3+4\frac{m_{\rm DM}^4}{m_{W(Z)}^4}-4
\frac{m_{\rm DM}^2}{m_{W(Z)}^2}\right),\nn\\
& &a_t= 24 \kappa_s^2\Delta_{h}^2  
m_t^2(m_{\rm DM}^2-m_t^2),~
a_h 
=\kappa_s^2\left(
1+24 \lambda_H \Delta_{h} \frac{m_W^2}{g^2}
\right)^2\nn
\ee
with $ \Delta_{h}=(4 m_{\rm DM}^2-m_h^2)^{-1}$
[$m_h$ is the corrected Higgs mass which should be compared with
$m_{h0}$ of (\ref{vev2}).]
The DM relic abundance is 
$\Omega \hat{h}^2 =(N_f^2-1)\times 
(Y_\infty s_0 m_{\rm DM})/(\rho_c/\hat{h}^2)$,
where $Y_\infty$ is the asymptotic value of the ratio 
$n_{\rm DM}/s$;
$s_0=2890/\mbox{cm}^3$ is the entropy density at present;
$\rho_c=3 H^2/8 \pi G=1.05 \times 10^{-5}\hat{h}^2 ~\mbox{GeV}/\mbox{cm}^3$ is the critical density;
$\hat{h}$ is the dimensionless Hubble parameter;
 $M_{pl}=1.22\times 10^{19}~ \mbox{GeV}$ is the  Planck energy;
and $g_*=106.75+N_f^2-1$ is the number of the effectively massless degrees of freedom
at the freeze-out temperature.
To obtain $Y_\infty$ we solve the Boltzmann equation
\be
\frac{d Y}{dx}=
-0.264~ g_*^{1/2} \left( \frac{m_{\rm DM} M_{\rm PL}}{x^2}\right) 
\langle v\sigma \rangle 
\left(  Y^2 -\bar{Y}^2\right)\nn
\ee
numerically, where $x$ is the inverse temperature $m_{\rm DM}/T$,
and $\bar{Y}$ is $Y$ in thermal equilibrium.
The spin-independent 
elastic cross section off the nucleon 
$\sigma_{SI}$
can be obtained from \cite{Barbieri:2006dq}
\be
\sigma_{SI}
&=&\frac{{1}}{4\pi} 
\left(\frac{\kappa_t\hat{f} m_N ^2}{m_{\rm DM}m_h^2}
\right)^2
\left(\frac{m_{\rm DM}}{m_N+m_{\rm DM}}
\right)^2,\nn
\ee
where $\kappa_t$ is given in
 (\ref{kappa}), $m_N$ is the nucleon mass, and
$\hat{f}\sim 0.3$ stems from the nucleonic matrix element 
\cite{Ellis:2000ds}.\footnote{If the value $\hat f$ improved by the recent lattice simulation~\cite{Oksuzian:2012rzb} is used, we obtain slightly smaller values (about $20\%$).}

%\begin{center}
\begin{figure}
 \includegraphics[width=12cm]{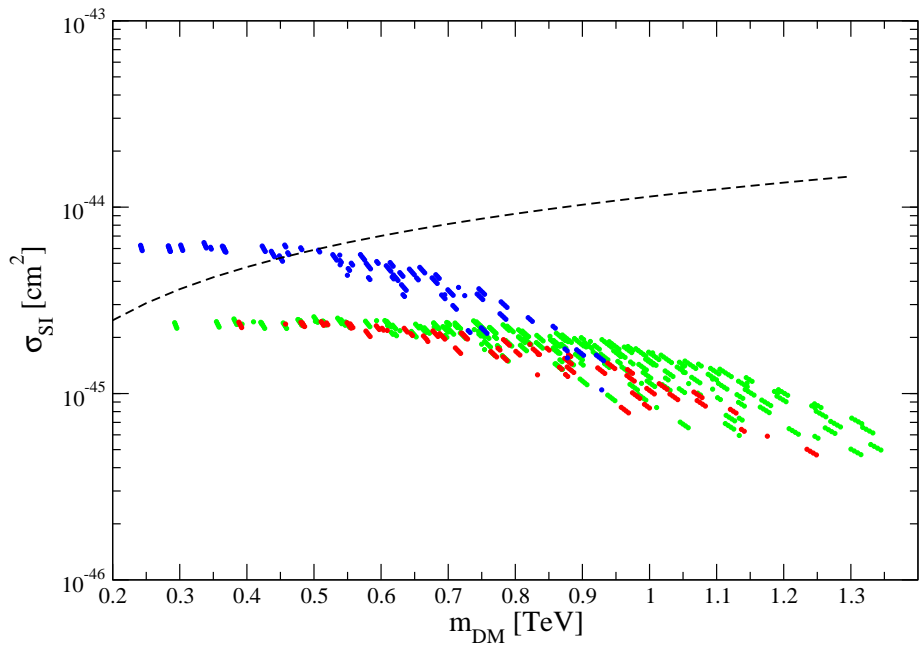}
\caption{\label{sigma}\footnotesize
The spin-independent elastic cross section $\sigma_{SI}$ 
of DM off   the nucleon as a function of $m_{\rm DM}$ for $N_f=2,N_c=5$ (red)
and $8$ (green)
and for $N_f=3,N_c=6$ (blue),
where $\Omega \hat{h}^2$  is required to be consistent with 
the PLANCK  experiment at $2\sigma$ level \cite{Planck:2015xua}.
The black dashed line stands for the central value of the LUX upper bound
\cite{Akerib:2013tjd}.}
\end{figure}
%\end{center}
Before we scan the parameter space, we consider a representative point
in the four-dimensional parameter space of the scalar couplings
with $N_f=2$ and $N_c=5$:
\be
\lambda_{S}&=& 1.20, \lambda'_{S}=5.38,~
 \lambda_{HS}=0.0525, \lambda_H=0.130,\nn
 \ee
which give
$f_0= 0.0749~\mbox{TeV}^2$, $M_0= 1.08~\mbox{TeV}$, $m_{\rm DM}= 0.801~\mbox{TeV}$, $m_\sigma =1.98~\mbox{TeV}$,\\
$\Lambda_H=0.501~\mbox{TeV}$, $\Omega \hat{h}^2 = 0.121$, $\sigma_{SI}=1.68
\times 10^{-45}~\mbox{cm}^{2}$, $\kappa_s=0.3988$, $\kappa_t=0.3089$.
In Fig.~\ref{sigma} we show in the  
$m_{\rm DM}\mbox{-}\sigma_{SI}$ plane
the predicted area  for various $N_f$ and $N_c$.
The predicted  values of  $\sigma_{SI}$   are just below
the LUX upper bound (black dashed line) 
\cite{Akerib:2013tjd}  and  can be tested by XENON1T, whose sensitivity is  $O(10^{-47})$
 $\mbox{cm}^2$ \cite{Aprile:2012zx,Aprile:2015uzo}.
 If we increase $N_f$, 
 we have to suppress  $Y_\infty$,
 because $\Omega  \hat{h}^2 \propto ( N_f^2-1)Y_\infty$, 
 which requires  a larger $\langle v \sigma\rangle$,
 leading to a larger $\sigma_{\rm SI}$.

\section{Summary}
We have assumed that the SM without the Higgs mass term 
is coupled through a Higgs portal term with a  
classically scale invariant  gauge sector,
which contains $N_f$ scalar fields.
Due to the strong confining force the 
gauge invariant scalar bilinear forms a  condensate,
thereby violating scale invariance.
The  Higgs portal term  is responsible for the transmission 
of the scale to the SM sector, realizing  electroweak scalegenesis.
We have formulated an effective  theory for 
the condensation of the scalar bilinear.
The excitation of the condensate is identified as DM,
where its scale is dynamically generated in the hidden gauge sector.
Our  formalism  is  simple and its application will be  multifold.
We have found that 
the DM mass is of $O(1)$ TeV and 
the predicted spin-independent elastic cross section off the nucleon 
is slightly below the LUX upper bound
and could be tested by  the XENON1T experiment.

\vspace{0.3cm}
\noindent{\bf Acknowledgements:} \\
We thank K.~S.~Lim,  M.~Lindner  and S.~Takeda for useful
discussions. We also thank the theory group of the Max-Planck-Institut
f\"ur  Kernphysik for their kind hospitality.
The work of M.~Y is supported by 
a Grant-in-Aid for JSPS Fellows (No. 25-5332).

\end{document}